\documentclass{aa}
\usepackage{natbib}
\bibpunct{(}{)}{;}{a}{}{,} % to follow the A&A style
\usepackage{amssymb}
\usepackage{graphicx}
\begin{document}
% Shortcuts for frequently used commands
\newcommand{\beq}{\begin{equation}}
\newcommand{\eeq}{\end{equation}}
\newcommand{\bea}{\begin{eqnarray}}
\newcommand{\eea}{\end{eqnarray}}
\newcommand{\bfig}{\begin{figure}[!t]}
\newcommand{\efig}{\end{figure}}
\newcommand{\bfigc}{\begin{figure}[!t]\begin{center}}
\newcommand{\efigc}{\end{center}\end{figure}}
\newcommand{\bt}{\begin{table}}
\newcommand{\et}{\end{table}}
\newcommand{\btu}{\begin{tabular}}
\newcommand{\etu}{\end{tabular}}
\newcommand{\bc}{\begin{center}}
\newcommand{\ec}{\end{center}}
\newcommand{\bi}{\begin{itemize}}
\newcommand{\ei}{\end{itemize}}
\newcommand{\bd}{\begin{description}}
\newcommand{\ed}{\end{description}}
\def\Rad{\tilde\omega}
\def\sol{ \ \mathrm{M}_\odot}
\newcommand{\tensor}[1]{\buildrel\leftrightarrow\over #1}
\title{On the Signatures of Gravitational Redshift: \\
The Onset of Relativistic Emission Lines}
\author{Andreas M\"uller\inst{1}
\and Margrethe Wold\inst{2}}
\titlerunning{Signatures of Gravitational Redshift}
\authorrunning{A. M\"uller \and M. Wold}
\institute{Max--Planck--Institut f\"ur Extraterrestrische Physik,  p.o. box 1312, D--85741 Garching, Germany
\and European Southern Observatory, Karl--Schwarzschild-Strasse 2, D--85748 Garching, Germany}
\offprints{Andreas M\"uller,
\email{amueller@mpe.mpg.de}}
\date{Received 16 May 2006 / Accepted 09 June 2006}
% acceptance paper with reference number -
%
%
% ABSTRACT
%
%
\abstract
% NEW at A&A: abstract in three parts: aims, methods, results are mandatory!
% context heading (optional)
{}
% aims heading (mandatory)
{We quantify the effect of gravitational redshift on emission lines to explore the 
transition region from the Newtonian to the Einsteinian regime. With the emitting 
region closer to the Kerr black hole, lines are successively subjected to a 
stronger gravitationally induced shift and distortion. Simulated lines are compared 
to broad, optical emission lines observed in Mrk~110.}
% methods heading (mandatory)
{We simulate relativistic emission line profiles by using Kerr ray tracing 
techniques. Emitting regions are assumed to be thin equatorial rings in stationary 
Keplerian rotation.
% inclined to 1 and 75 degrees and ranging from 2 to $10^5$ 
% gravitational radii. 
The emission lines are characterised by a generalized Doppler factor or redshift 
associated with the line core.}
% results heading (mandatory)
{With decreasing distance from the black hole, the gravitational redshift 
starts to smoothly deviate from the Newtonian Doppler factor: 
% of unity 
% (face--on): 
Shifts of the line cores reveal an effect at levels of 0.0015 to 60~\% at 
gravitational radii ranging from $10^{5}$ to 2. This corresponds to fully 
relativistic Doppler factors of 0.999985 to 0.4048. 
% TODO: check this trafo
% but it better in conclusions of abstract
% For the H$\beta$ line this corresponds to a shift of the line center of 5--10 km/s.
The intrinsic line shape distortion by strong gravity i.e.\ very asymmetric 
lines occur at radii smaller than roughly ten gravitational radii.}
% conclusions heading (optional)
{Due to the asymptotical flatness of black hole space--time, GR effects are 
ubiquitous and their onset can be tested observationally with sufficient 
spectral resolution. 
With a resolving power of $\sim100000$, yielding a
resolution of $\approx0.1$ {\AA} for optical and near--infrared 
broad emission lines like H$\beta$, HeII and Pa$\alpha$, the gravitational
redshift can be probed out to approximately 75000 gravitational radii. 
% by instruments like UVES and CRIRES on the VLT. 
In general, gravitational redshift is an important indicator of black hole mass and 
disk inclination as recently demonstrated by observations of optical lines in Mrk~110. 
Comparing our simulated lines with this observations, we independently confirm an 
inclination angle of 30 degrees for the accretion disk. Redshift deviations induced by 
black hole spin can be probed only very close to the black hole e.g. with X--ray iron 
lines.
}
\keywords{Black hole physics -- Relativity --  Line: profiles -- Galaxies: active -- Galaxy: nucleus -- Galaxies: Seyfert}
\maketitle
%
% INTRO
%
%
\section{Introduction}
Active galactic nuclei (AGN) such as Seyfert galaxies and quasars are powered 
by accreting supermassive black holes (SMBHs) following the standard model 
that has been developed over four decades \citep{Lynden-Bell1969, Lynden-Bell1971}. 
The masses of SMBHs lie in the range $10^{6}$ to $10^{10}$ M$_{\odot}$, see e.g 
\cite{Netzer2003}. 
In the standard model, clouds moving in the gravitational potential of the 
black hole are photoionized by the central AGN continuum, thereby producing
Doppler broadened emission lines with widths of typically 10$^{3}$--10$^4$ 
km\,s$^{-1}$ \citep[~as pioneering studies]{Woltjer1959}. The region where the 
broad lines originate is usually referred to as the broad--line region (BLR). The 
scale of the BLR is believed to be $10^{15}$ to $10^{17}$ cm, corresponding to 
$\sim 10^{3}$ to $\sim 10^{5} \ r_\mathrm{g}$ or 0.6 to 60 light days for a 
10$^{7}$ M$_{\odot}$ black hole. Here the gravitational radius is defined as 
$r_{\rm g}= GM/c^{2}$ with Newton's constant ${\rm G}$, vacuum speed of light 
${\rm c}$ and black hole mass $M$. \cite{Robinson1990} have presented complex models 
involving spherical or disk geometries for the BLR as well as rotational and radial 
cloud kinematics. They studied the influence of continuum variability on line 
profiles and found diverse line shapes exhibiting spikes, bumps and shoulders 
though in a non--relativistic regime. Additional velocity components in the BLR caused 
by disk winds \citep{Konigl1994} and a radial component have been suggested from accretion 
theory and radial velocity maps of the narrow--line region \citep{Ruiz2001}. However, 
the detailed structure and velocity field of the BLR remain unclear \citep{Collin2006}. \\
The broad--line clouds respond to variations in the central photoionizing continuum as 
suggested by strong correlations between H$\beta$ response times and non--stellar optical
continuum fluxes, $\tau_\mathrm{cent}\propto\sqrt{F_\mathrm{UV}}$ \citep{Peterson2002}. 
This phenomenon is exploited in reverberation mapping techniques to determine both the 
scale of the BLR and the black hole mass \citep{Blandford1982, Peterson1993, Kaspi2000}. \\
The idea that gravitational redshift may influence optical lines causing line asymmetries 
was raised by \citet{Netzer1977}. In the Seyfert--1 galaxy Akn~120, a slight redward 
displacement of the H$\beta$ line was reported, amounting to $\Delta z\sim 0.0013$, 
interpreted as the result of gravitational redshift \citep{Peterson1985}. 
However, such effects may also arise from attenuation of the BLR or light--travel 
time effects, as discussed by \cite{Peterson1985}. Similar studies that assume that observed
effects are a result of gravitational redshift have been done for a quasar sample where the 
SMBH mass of QSO~0026+129 could be roughly estimated to be $2\times 10^9\sol$ \citep{Zheng1990}; 
this is still the current value within a factor of 2 \citep{Czerny2004}. Recently, several BLR 
optical emission lines in the narrow--line Seyfert--1 galaxy Mrk~110 were investigated
\citep[see][ ~K03 hereafter]{Kollatschny2003}. 
In that work, H$\alpha$, H$\beta$, HeI$\lambda$5876 and HeII$\lambda$4686 emission lines were 
found to possess a systematic shift to the red, with higher ionization lines showing larger shifts 
as expected in a BLR with stratified ionization structure. \\
In this paper, we study the gravitational redshift over a large range of distances from the central 
black hole. We quantify the relativistic gravitational redshift on emission lines until GR fades 
beyond the current observable limit. The investigation is carried out in a very general form by 
discussing the observed line profile as a function of the generalized GR Doppler factor ($g$--factor) 
for Kerr black holes and an arbitrary velocity field of emitters, see e.g.\ \citet[ ~M04 hereafter]{Mueller2004}. 
Pioneering work on relativistic spectra was performed by \citet{Cunningham1975} using transfer functions. 
However, the considerations of the $g$--factors in this work were restricted to minimum and maximum values 
of $g$ on infinitesimally narrow and thin stationary rings. Furthermore, the distance range of interest 
for BLRs, $10^{3}$ to $10^{5} \ r_\mathrm{g}$, has not been investigated in detail. \\
\cite{Corbin1997} studied relativistic effects on emission lines from the BLR by assuming Keplerian 
orbits for the emitting clouds in a Schwarzschild geometry. It was found that line profiles decrease 
in both, width and redward centroid shift when the line emitting region moves away from the black hole. \\
% He was however not able to show that the redward line displacement due to gravitational redshift decreases 
% as the line emitting region moves away from the black hole. 
% More precisely, it has been reported that then line profiles decrease in both, 
% width and redward centroid shift. \\
Our goal is to accurately quantify the effects of gravitational redshift in the vicinity of a Kerr black 
hole. After a very general consideration that holds for any classical black hole of arbitrary mass, a more 
specific treatment involving optical emission lines from BLRs is addressed. For the case study of Mrk~110, 
it is even demonstrated how the mass of the SMBH and the inclination of the inner disk can be determined.
\section{Method, analysis tools and model}
\subsection{Relativistic ray tracing} \label{sec:raytracer}
In contrast to Cunningham's work, emission lines are computed by ray tracing in the Kerr geometry 
of rotating black holes. Light rays emitted in the vicinity of the black hole travel to the 
observer on null geodesics in curved space--time, and in this work the observer is assumed to be 
located at $r_\mathrm{obs}=10^{7} \ r_\mathrm{g}$. The \textit{Kerr Black Hole Ray Tracer} (KBHRT) maps 
emitting points in the equatorial plane of a Kerr black hole to points on the observer's screen. 
Spectral line fluxes are computed by numerical integration over the solid angle subtended by the screen. 
All relativistic effects such as gravitational redshift, beaming and lensing are included, but higher order 
images are not considered. The complete solver has been presented in earlier work (M04).

\subsection{Analysing relativistic emission lines} \label{sec:AnaLines}
In the following, line fluxes are discussed as a function of the 
$g$-factor which is defined as
\beq \label{eq:g}
g=\nu_\mathrm{obs}/\nu_\mathrm{em}=\lambda_\mathrm{em}/\lambda_\mathrm{obs}=\frac{1}{1+z},
\eeq
where $\nu$ and $\lambda$ denote frequency and wavelength, respectively, and the redshift is $z$. 
Emitter's and observer's frame of reference are indicated by subscripts 'em' and 'obs'. A $g$-factor 
of unity therefore corresponds to an unshifted line, whereas $g < 1$ indicates redshifted emission 
and $g > 1$ blueshifted emission. Note that by using the $g$-factor, the simulated line profiles are 
discussed very generally \textit{without} specifying a particular emission line. 

Analysis tools for relativistic emission lines as well as line classification schemes 
by morphology that were introduced by M04 (Sec. 8) are utilized here. A relativistic 
emission line exhibiting two Doppler peaks can be characterised by several quantities:
\bi
\item $g_\mathrm{min}$, minimum $g$--factor that defines the terminating energy at the red wing
\item $g_\mathrm{max}$, maximum $g$--factor that defines the terminating energy at the blue wing
\item $g_\mathrm{rp}$, the $g$--factor associated with the red relic Doppler peak
\item $g_\mathrm{bp}$, the $g$--factor associated with the blue beamed Doppler peak
\item DPS = $g_\mathrm{bp}$ - $g_\mathrm{rp}$, the Doppler peak spacing in energy 
\item $F_\mathrm{max}$, maximum line flux
\item $F_\mathrm{rp}$, line flux associated with the red relic Doppler peak
\item $F_\mathrm{bp}$, line flux associated with the blue beamed Doppler peak
\item DPR = $F_\mathrm{bp}$/$F_\mathrm{rp}$, the flux ratio of both Doppler peaks
\ei
The existence and strength of these parameters depend on the line shape. Low inclination angles 
of the emitting surface i.e.\ face--on situations with axial observers destroy the typical 
structure with two distinct Doppler peaks.

Based on relativistic emission line terminology, line morphologies can be classified as 
triangular, double--peaked, double--horned, shoulder--like and bumpy shapes. Triangular 
and shoulder--like morphologies lack a red Doppler peak. Bumpy morphology even lacks a 
distinct blue beaming peak either because a very steep disk emissivity suppresses emission 
at large disk radii or because the line originates too close to the black hole. We want to 
stress here that \cite{Robinson1990} found a similar terminology for non--relativistic lines 
but the classification scheme for relativistic lines in M04 was established independently. 
% definition of line core energy -> g_core

%A new quantity is presented here:
In order to be able to characterize any line profile independently of its
morphology, and in order quantify the shift of the resulting line centroid, 
we define a new quantity, $g_{\mathrm core}$:
\beq \label{eq:gcore}
g_\mathrm{core}=\frac{\sum_i \ g_\mathrm{i}\,F_\mathrm{i}}{\sum_i \ F_\mathrm{i}}.
\eeq
The $g_{\mathrm core}$ parameter is thus associated with the line core energy, 
i.e.\ the energy associated with the flux weight of the whole line. As can be
seen, it is evaluated numerically by multiplying each energy bin, $g_\mathrm{i}$, 
with the corresponding flux in that bin, $F_\mathrm{i}$, and summing over the line 
profile. 

Gravitational redshift in the weak field regime establishes pure shifts of spectral 
features without changing their intrinsical shape. However, gravitational redshift
in the strong field regime -- \textit{strong gravity} -- produces remarkable distortions
of spectral shapes if the rest frame feature is compared to its analogue in the observer's 
frame. Distortion is a key feature of relativistic spectra exhibiting very skewed and 
asymmetric line profiles \citep{Fabian1989, Popovic1995, Tanaka1995}. These effects are 
important only very close to the black hole. Here we investigate how gravitational redshift 
changes the mode i.e.\ Einsteinian gravity transmutes to Newtonian gravity for emission 
regions moved away from the black hole. 

We also define the \textit{half--energy radius} as the radius where a given $g$--factor 
is reduced to exactly $1/2$ its original value. More precisely, $R_j$ denotes the half--energy 
radius associated with $j\in\{g_\mathrm{min},\, g_\mathrm{max},\, g_\mathrm{core},\, g_\mathrm{rp},\, g_\mathrm{bp}\}$. 
Due to Eq. (\ref{eq:g}) the observed energy of the radiation associated with the specific 
$g$ is exactly one half of the emitted energy in the rest frame. As a measure of strong 
gravity $R_\mathrm{grp}$ is chosen as the radius where the $g$--factor associated with the 
red Doppler peak is exactly $1/2$. This is a suitable choice because strong gravity 
deforms the red line wing in an extraordinary manner.
\subsection{Emitter model: Rendering parameters and emissivity} \label{sec:param}
The aim of this paper is to study gravitational redshift effects. Therefore, it is of particular interest 
to avoid blueshift effects that would blur or cancel the redshift. An easy way to switch off blueward displacements 
is tilting the emitting region to a face--on situation: axial observers with inclination angle $i=0^\circ$ 
to the emitting area have no relative motion along the line of sight to the emitter. In this case, the 
radiation is only affected by gravitational redshift because no motion is directed out of the plane. 

For numerical reasons an inclination angle of $i=1^\circ$ is chosen which is close enough to the 
face--on situation. Other parameters are: Kerr black hole rotating at Thorne's limit $a/M=0.998$ 
\citep{Thorne1974} and a prograde Keplerian velocity field of emitters, 
$\Omega_\mathrm{K}=\sqrt{M}/(\sqrt{r^3}+a\sqrt{M})$. We consider here stationary thin rings with 
emission peaking at $R_\mathrm{peak}$. This is established by rendering a disk and shifting a 
Gaussian radial emissivity profile over the disk 
\beq
\epsilon(r)\propto\exp\left(-\frac{(r-R_\mathrm{peak})^2}{\sigma_\mathrm{r}^2}\right),
\eeq
as introduced by M04. The parameter $\sigma_\mathrm{r}$ controls the width of the Gaussian or the size of the 
emitting region and is chosen to be 0.2. The peak radius, $R_\mathrm{peak}$, is given in units of 
gravitational radius. A localized Gaussian emissivity set in this way mimics a thin and narrow 
luminous ring with $\simeq 1 \ \mathrm{r_g}$ distance between inner and outer edge.  The Gaussian guarantees
a smooth but steep decay of emission at the ring edges. Note that using $\mathrm{r_g}$ as a natural scale allows 
for applications to any classical black hole including also stellar and intermediate--mass black holes. 

The technique is now straightforward: The ring is shifted from large distances in asymptotically flat space--time 
in the direction towards the black hole where space-time curvature becomes strong. In the present work, we assume 
that the radial range of interest is $2 \ \mathrm{r_g} < R_\mathrm{peak} < 100000 \ \mathrm{r_g}$. For each simulated 
emission ring we determine the line core energy $g_\mathrm{core}$ by applying Eq. (\ref{eq:gcore}) and computing 
the associated line redshift, $z_\mathrm{core}$, from Eq. (\ref{eq:g}). These quantities allow for quantifying the 
gravitational redshift as a function of distance to the rotating black hole. 

% now coming to BLR
Further, we use the ring emitter model as a simple model for the broad--line region in AGN. In this case, the BLR 
clouds are distributed in an equatorial plane and follow Keplerian orbits around the black hole as proposed elsewhere 
\citep{Netzer1977, Robinson1990, Corbin1997}. Depending on radial distance to the center and orientation angle to 
observer, the BLR emission is influenced by Doppler effect, gravitational redshift and beaming to different extent. 
We show in Sec. \ref{sec:gropt} that it is possible to fit optical data to this simple, flat Keplerian BLR model and 
that plausible inclination angles of the inner disk can be deduced.
\section{Gravitationally redshifted emission lines} \label{sec:GRlines}
\subsection{Redshifted line cores} \label{sec:linecore}
%
%%%%%%%%%%%%%%%%%%%%%%%%%%%%%%%%%%%%%%%%%%%%%%%%%%%%%%%%%%%%%%%%%%%%%%%%%%%%%%%%%%%%%%%%%%%%%%%%%%
%
% Fig. 1: Gravitational redshift of line cores depending on position of emitter
\bfigc
	\rotatebox{0}{\includegraphics[width=0.5\textwidth]{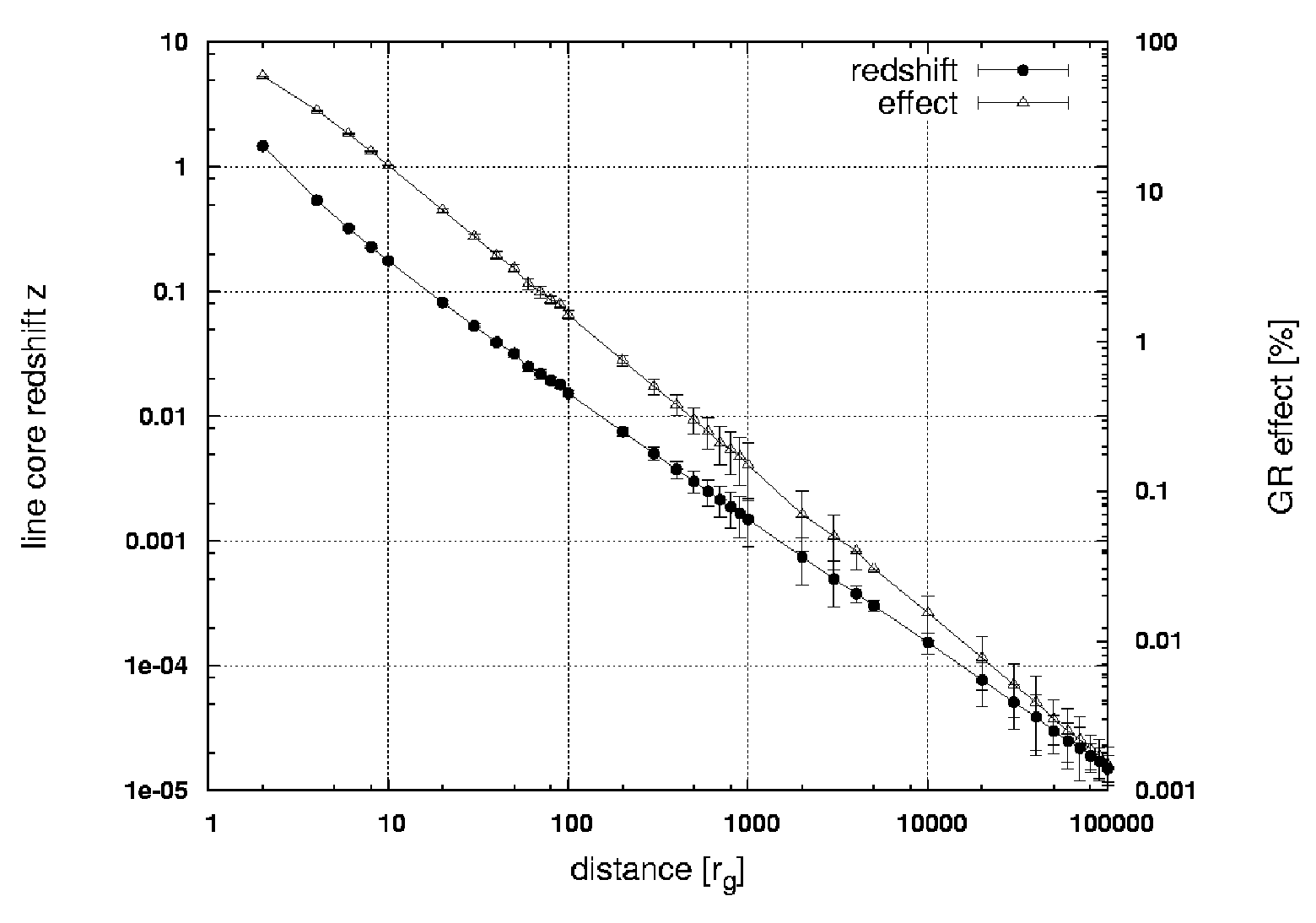}}
	\caption{Radial dependence of redshift $z$ of line cores (\textit{filled circles, 
	left y--axis}) and strength of gravitational redshift effect (\textit{triangles, 
	right y--axis}). Inclination angle amounts $1^\circ$.} \label{fig:z-r-plot}
\efigc
%%%%%%%%%%%%%%%%%%%%%%%%%%%%%%%%%%%%%%%%%%%%%%%%%%%%%%%%%%%%%%%%%%%%%%%%%%%%%%%%%%%%%%%%%%%%%%%%%%
As outlined in the previous section the line cores are computed by ray tracing and their dependence on 
radial distance to the black hole is analysed. Fig. \ref{fig:z-r-plot} displays the core redshifts of 
emission lines, $z_\mathrm{core}$, as a function of $R_\mathrm{peak}$ for rings inclined to $i=1^\circ$. 
The redshifted lines are dominated by gravitational redshift by construction. The filled circles show 
that the redshift approaches $z\rightarrow 0$, i.e.\ that $g\rightarrow 1$, at distances of a few 
thousand gravitational radii from the black hole. This is the regime of nearly flat space--time and 
Newtonian physics. But approaching the black hole, space--time curvature becomes more significant: 
$z$ grows rapidly and $g$ approaches zero. 

The triangles illustrate the strength of the gravitational redshift effect in an alternative way. 
The scale of the axis on the right--hand side is computed as $(1-g_\mathrm{core})\times\,100$ so 
that a value of $g=1$ corresponds to 0\% effect and $g=0$ to a 100\% effect, i.e.\ the gravitational 
redshift at the event horizon of the black hole. The error bars result from uncertainties in determining 
$g$--factors. Ray tracing simulations with different numerical resolutions in both disk resolution for 
rendering and spectral resolution for line computation yield slightly deviating values for $g_\mathrm{core}$ 
(and other quantities in general). 

Gravitational redshift can alternatively be visualised by plotting the line core energies, $g_\mathrm{core}$, 
as a function of the peak radius for each ring. Fig. \ref{fig:gcore-i1vs75} shows the result for $i=1^\circ$ 
and $i=75^\circ$. Highly inclined rings exhibit strong blueshift effects overlapping the redshift. As a consequence 
the core $g$--factors of the $i=75^\circ$ dataset never drop below 0.8 for radii larger than the marginally 
stable orbit. So, comparison of both orientations demonstrates that highly inclined rings are not appropriate 
to study gravitational redshift as pure shifting effect. It is shown later that high inclinations are 
well--suited for probing strong gravity. 

In Tab. \ref{tab:01} we show the results of the Kerr ray tracing simulations for nearly face--on rings, 
$i=1^\circ$, in terms of numerical values for $g_\mathrm{core}$, $z_\mathrm{core}$ and GR effect as
a function of distance to a Kerr black hole with $a/M=0.998$. 
\bt
\caption{Radial redshift dependence for $i=1^\circ$}
\label{tab:01}
\centering
\btu{cccc}
\hline
\hline
  &  &  & \\
$R_\mathrm{peak}$ [$r_\mathrm{g}$] & $g_\mathrm{core}$ & $z_\mathrm{core}$ & GR effect [\%] \\
  &  &  & \\
\hline
\hline
2 & 0.404763 & 1.470584 & 59.52 \\
4 & 0.649680 & 0.539220 & 35.03 \\
6 & 0.755815 & 0.323075 & 24.42 \\
8 & 0.813688 & 0.228973 & 18.63 \\
10 & 0.849776 & 0.176780 & 15.02 \\
\hline
20 & 0.924283 & 0.081919 & 7.57 \\
30 & 0.949551 & 0.053130 & 5.04 \\
40 & 0.962198 & 0.039287 & 3.78 \\
50 & 0.969071 & 0.031916 & 3.09 \\
60 & 0.975612 & 0.024997 & 2.44 \\
70 & 0.978597 & 0.021871 & 2.14 \\
80 & 0.980951 & 0.019419 & 1.90 \\
90 & 0.982324 & 0.017994 & 1.77 \\
100 & 0.984930 & 0.015301 & 1.51 \\
\hline
200 & 0.992485 & 0.007572 & 0.75 \\
300 & 0.994996 & 0.005029 & 0.50 \\
400 & 0.996250 & 0.003764 & 0.38 \\
500 & 0.997002 & 0.003007 & 0.30 \\
600 & 0.997503 & 0.002504 & 0.25 \\
700 & 0.997860 & 0.002144 & 0.21 \\
800 & 0.998128 & 0.001875 & 0.19 \\
900 & 0.998337 & 0.001666 & 0.17 \\
1000 & 0.998503 & 0.001499 & 0.15 \\
\hline
2000 & 0.999253 & 0.000747 & 0.07 \\
3000 & 0.999502 & 0.000498 & 0.05 \\
4000 & 0.999621 & 0.000380 & 0.04 \\
5000 & 0.999696 & 0.000304 & 0.0304 \\
10000 & 0.999846 & 0.000154 & 0.0154 \\
\hline
20000 & 0.999923 & 0.000077 & 0.0077 \\
30000 & 0.999949 & 0.000051 & 0.0051 \\
40000 & 0.999961 & 0.000039 & 0.0039 \\
50000 & 0.999970 & 0.000030 & 0.0030 \\
60000 & 0.999975 & 0.000025 & 0.0025 \\
70000 & 0.999978 & 0.000022 & 0.0022 \\
80000 & 0.999981 & 0.000019 & 0.0019 \\
90000 & 0.999983 & 0.000017 & 0.0017 \\
100000 & 0.999985 & 0.000015 & 0.0015 \\
\hline
\hline
\etu
\et
In principle, Tab. \ref{tab:01} illustrates the die out of GR with increasing radius. However, it is important 
to note that according to the asymptotical flatness of GR black holes solutions, space--time curvature approaches 
zero only in the limit $r\rightarrow\infty$ i.e.\ there is no finite distance at which the gravitational redshift 
vanishes exactly. Hence, Tab. \ref{tab:01} could be generally continued \textit{ad infinitum}. However, 
observability poses a limit to what is practical since the resolving power of a spectrograph, in the ideal case, 
constrains the amount of gravitational redshift (parameterized by $z_\mathrm{core}$ here) that can be detected. 
Assuming a spectral resolution of 0.1 \AA\, for H$\beta$ as is obtainable by instruments like UVES and CRIRES on 
the VLT, the corresponding critical value of the $g$-factor is $g=0.999979$. In terms of velocity shift, this is 
$\approx10$ km\,s$^{-1}$. As seen in Tab. \ref{tab:01}, this shift occurs at a radius of $\sim 75000 \ r_\mathrm{g}$. 
Hence, we do not consider radii above $100000 \ r_\mathrm{g}$ in Tab. \ref{tab:01}. For supermassive black holes of 
10$^{7}$--10$^{8}$ M$_{\odot}$ this radius corresponds to 0.05--0.5 pc, whereas for stellar--mass black holes of 
$\sim 10$ M$_{\odot}$ it is 0.01~AU.

Applying Eq. (\ref{eq:g}) 
% g = lambda_em/lambda_obs
an optical HeII emission line with $\lambda_\mathrm{em}=4686$ \AA\, in the emitter frame is gravitationally 
redshifted to $4686.1,\,4686.7,\,4693.0,\,4757.7$ \AA \ at $100000,\,10000,\,1000,\,100 \ r_\mathrm{g}$.
% values computed from table g_cores: 4686.1 (100000), 4686.7 (10000), 4693.0 (1000), 4757.7 (100)
%
%%%%%%%%%%%%%%%%%%%%%%%%%%%%%%%%%%%%%%%%%%%%%%%%%%%%%%%%%%%%%%%%%%%%%%%%%%%%%%%%%%%%%%%%%%%%%%%%%%
%
% Fig. 2: g_core high i vs. low i
\bfigc
	\rotatebox{0}{\includegraphics[width=0.5\textwidth]{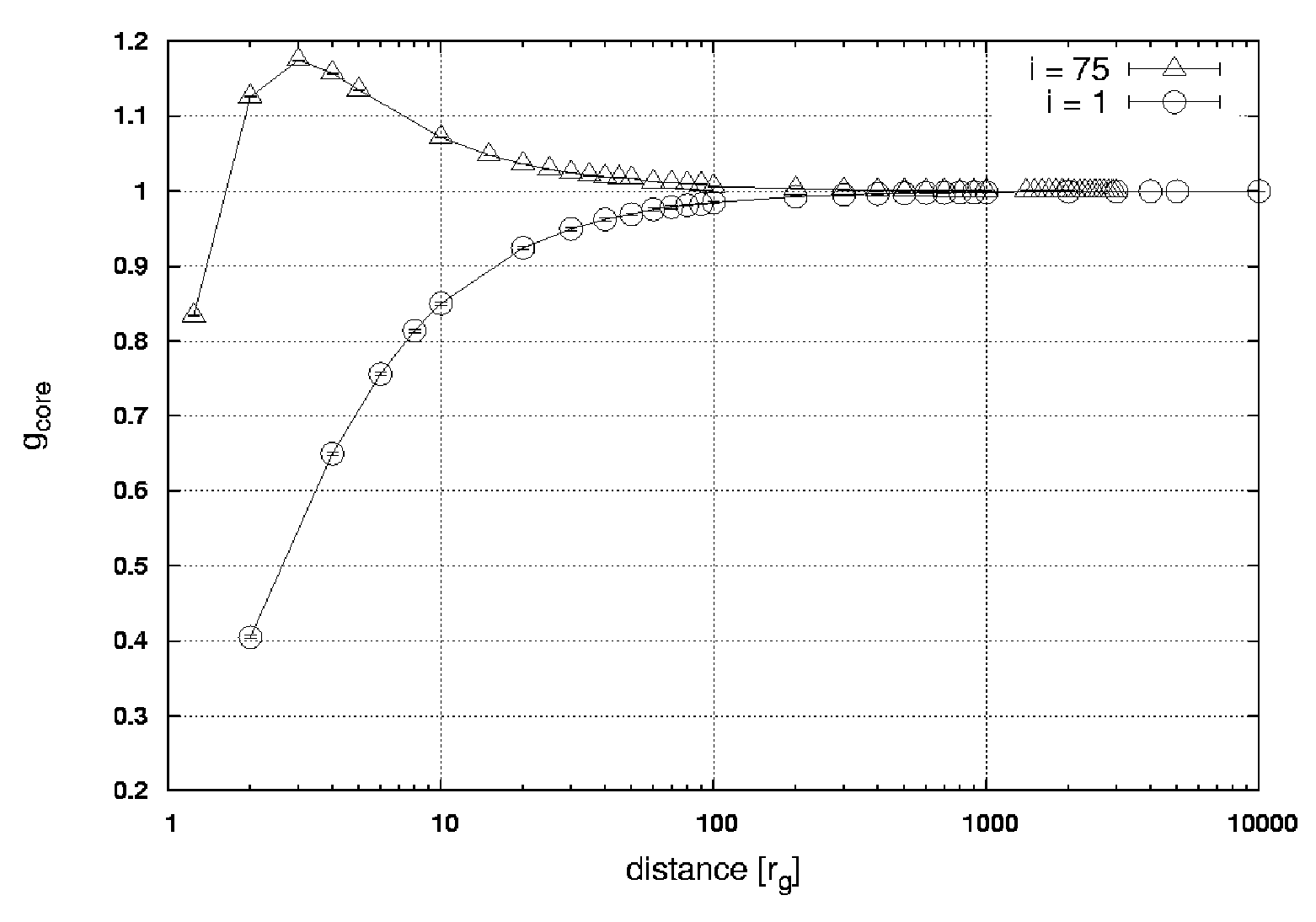}}
	\caption{Shift of line core energy in units of $g$--factor with distance to the black 
	hole. Lowly inclined rings, $i=1^\circ$, are compared to highly inclined rings, 
	$i=75^\circ$. Note the blueshift $g > 1$ in the latter case.} \label{fig:gcore-i1vs75}
\efigc
%%%%%%%%%%%%%%%%%%%%%%%%%%%%%%%%%%%%%%%%%%%%%%%%%%%%%%%%%%%%%%%%%%%%%%%%%%%%%%%%%%%%%%%%%%%%%%%%%%
%
\subsection{Line distortion by strong gravity} \label{sec:strong}
In the previous section the gravitational redshift was investigated as an effect that shifts the 
whole line core. We now focus on the behaviour of the intrinsic shape of relativistic 
emission lines as a function of distance to the black hole. This is dubbed strong gravity as 
anticipated in Sec. \ref{sec:AnaLines}. In the following, we tilt the emitting ring to an inclination 
of $i=75^\circ$ in order to be able to study the characteristic broad profile with the two relic 
Doppler peaks. Fig. \ref{fig:lines-i1+75} shows how the line profile changes by tilting from 
$i=1^\circ$ to $i=75^\circ$. 

Fig. \ref{fig:grav-i75} shows the effect of gravitational redshift on the $i=75^\circ$ profile 
as a function of distance to the black hole. The line profile broadens and the line flux gets more and 
more suppressed as the emitting ring is moved closer to the black hole\footnote{However, it must be noted 
that the line flux also gets suppressed because the effective emitting area decreases for smaller rings that 
have constant $r_\mathrm{out}-r_\mathrm{in}$ in any case.}. Eventually, the line decays and disappears at 
the event horizon. The red relic Doppler peak is shifted to lower energies as the ring approaches the 
hole, illustrated in 
% A practical way to illustrate this is to plot the radial dependence of the spectral position of the 
%red relic Doppler peak, $g_\mathrm{rp}$. 
Fig. \ref{fig:g_rp-i75}.
% is also based on Kerr ray tracing data 
%and illustrates how the red relic Doppler horn moves smoothly to lower values of $g_\mathrm{rp}$ i.e. to 
%lower energies. 
The red peak flux is also more and more suppressed as can be seen in Fig. \ref{fig:grav-i75}. Close to 
the black hole the distortions are so strong that the red Doppler peak becomes highly blurred and effectively 
vanishes in the line profile. At the event horizon the profile dies out and becomes unobservable.
%%%%%%%%%%%%%%%%%%%%%%%%%%%%%%%%%%%%%%%%%%%%%%%%%%%%%%%%%%%%%%%%%%%%%%%%%%%%%%%%%%%%%%%%%%%%%%%%%%
%
% Fig. 3: example line with g_core explanation
\bfigc
	\rotatebox{0}{\includegraphics[width=0.5\textwidth]{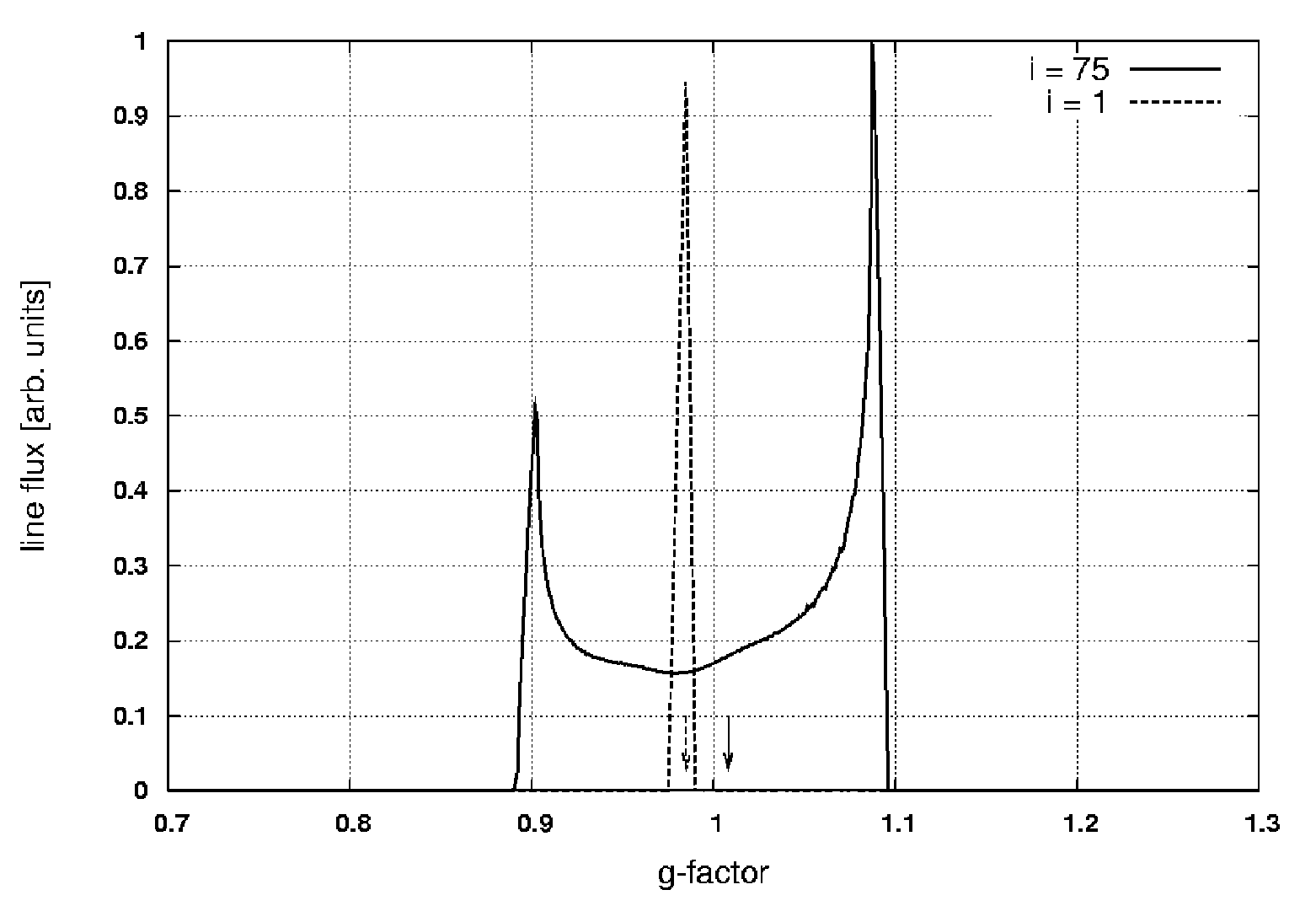}}
	\caption{Two example relativistic lines emitted from a highly inclined 
	ring, $i=75^\circ$, and a face--on ring, $i=1^\circ$, both with the 
	emission peaking at the same $R_\mathrm{peak}=100.0 \ r_\mathrm{g}$ 
	around a Kerr black hole with $a/M=0.998$. The arrows indicate the 
	core $g$--factors, $g_\mathrm{core}\simeq 1.008$ and $g_\mathrm{core}\simeq 0.985$ 
	respectively for each particular case.} \label{fig:lines-i1+75}
\efigc
%%%%%%%%%%%%%%%%%%%%%%%%%%%%%%%%%%%%%%%%%%%%%%%%%%%%%%%%%%%%%%%%%%%%%%%%%%%%%%%%%%%%%%%%%%%%%%%%%%
%%%%%%%%%%%%%%%%%%%%%%%%%%%%%%%%%%%%%%%%%%%%%%%%%%%%%%%%%%%%%%%%%%%%%%%%%%%%%%%%%%%%%%%%%%%%%%%%%%
%
% Fig. 4: Line distortion by strong gravity
\bfigc
 	\rotatebox{0}{\includegraphics[width=0.5\textwidth]{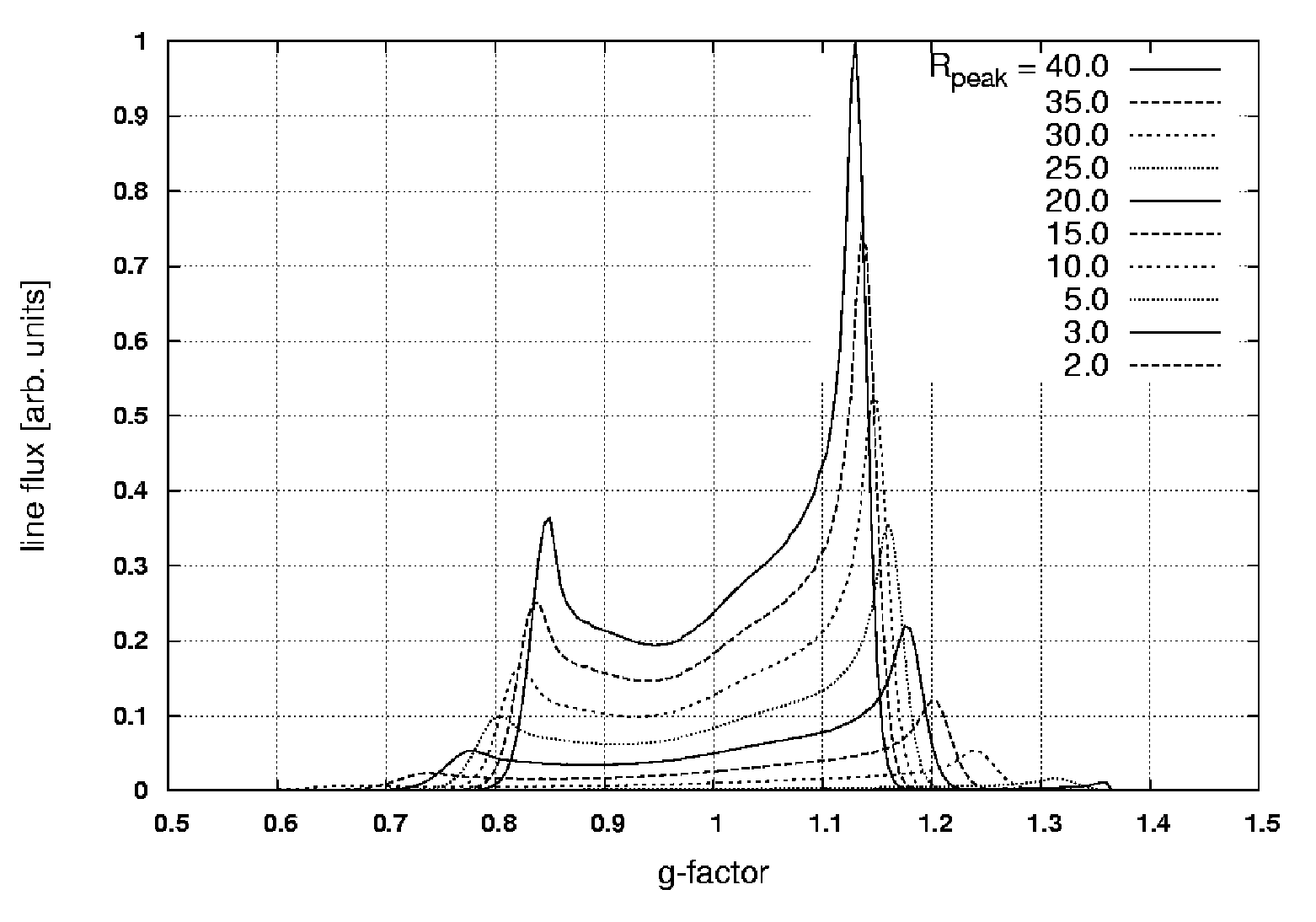}}
	\caption{Line distortion by strong gravity. The emission of narrow rings peaks at the
	radii as denoted in the legend. There is no way to confuse the lines because line 
	flux successively decreases with decreasing radius of maximum emission $R_\mathrm{peak}$. 
	All rings follow Keplerian rotation around a Kerr black hole with $a/M=0.998$ and all are 
	inclined to an inclination angle of $i=75^\circ$.} \label{fig:grav-i75}
\efigc
%%%%%%%%%%%%%%%%%%%%%%%%%%%%%%%%%%%%%%%%%%%%%%%%%%%%%%%%%%%%%%%%%%%%%%%%%%%%%%%%%%%%%%%%%%%%%%%%%%
%
Fig. \ref{fig:g_rp-i75} can be used to read the half--energy radius associated with the red relic Doppler 
peak as defined in Sec. \ref{sec:AnaLines}. At $i=75^\circ$ this value is $R_\mathrm{grp}\simeq 5 \ \mathrm{r_g}$ 
and documents that strong gravity is important only very close to the black hole. The distortion of the 
intrinsic line profile by gravitational redshift can be seen from $g_\mathrm{rp}$ alone or by using another 
line criterion. A further line characteristic is the Doppler peak spacing (DPS) i.e.\ the energetic distance 
of both Doppler peaks (if available). DPS can be measured in units of energy, frequency, wavelength or -- as 
has been done here for generality -- in units of $g$. Fig. \ref{fig:DPS-i75} shows that this quantity does not 
remain constant as the black hole is approached. DPS rises quickly so that the line is stretched. This 
phenomenon is directly but only qualitatively visible in Fig. \ref{fig:grav-i75}. Gravitational redshift causes 
an additional suppression in flux so that close to the black hole the emission line from an intermediately to 
highly inclined ring is asymmetric and skewed. 

We close this section with a comment on black hole detectability: Redshifted line cores as presented in 
Sec. \ref{sec:linecore} leave enough room for other gravitational sources than black holes; in contrast, 
spectral lines distorted by strong gravity in connection with a measured high compact mass support black hole 
candidates.
%% --------------------------------------------------------------------------------------------------------------   
%%%%%%%%%%%%%%%%%%%%%%%%%%%%%%%%%%%%%%%%%%%%%%%%%%%%%%%%%%%%%%%%%%%%%%%%%%%%%%%%%%%%%%%%%%%%%%%%%%
%
% Fig. 5: Position of red relic Doppler peak g_rp at 75 degree inclination
\bfigc
	\rotatebox{0}{\includegraphics[width=0.5\textwidth]{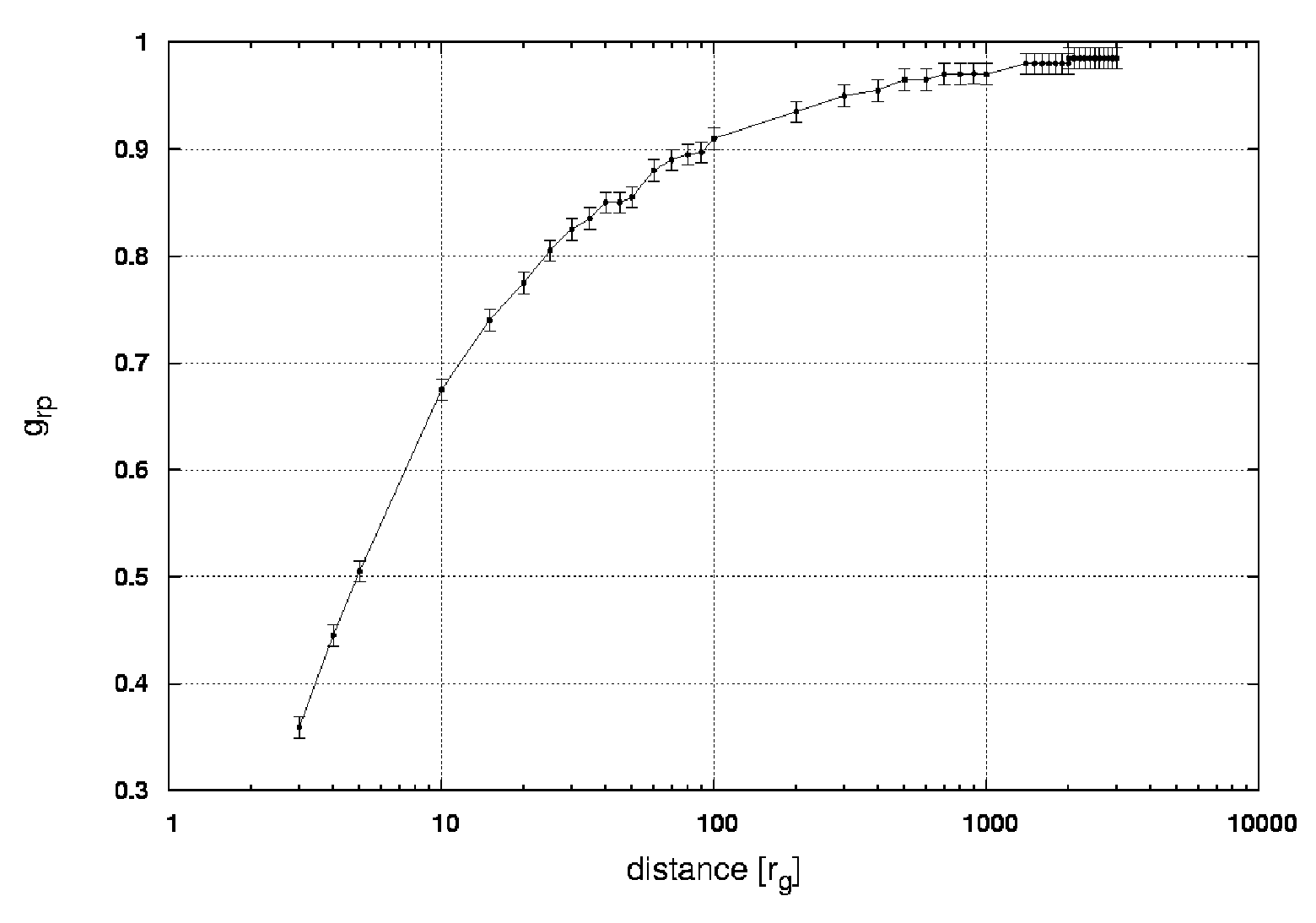}}
 	\caption{Distortion of the red relic Doppler peak for rings satisfying $i=75^\circ$: With 
	decreasing radius the red peak can be found at lower peak energies due to strong gravitational 
	redshift. Very close to the black hole -- $r\lesssim 3 \ \mathrm{r_g}$ at this specific 
	inclination -- these distortions are such strong that the red Doppler peak is highly blurred 
	and vanishes effectively in the line profile.} \label{fig:g_rp-i75} 
\efigc
%%%%%%%%%%%%%%%%%%%%%%%%%%%%%%%%%%%%%%%%%%%%%%%%%%%%%%%%%%%%%%%%%%%%%%%%%%%%%%%%%%%%%%%%%%%%%%%%%%
%%%%%%%%%%%%%%%%%%%%%%%%%%%%%%%%%%%%%%%%%%%%%%%%%%%%%%%%%%%%%%%%%%%%%%%%%%%%%%%%%%%%%%%%%%%%%%%%%%
%
% Fig. 6: Doppler peak spacing in units of g (DPS)
\bfigc
	\rotatebox{0}{\includegraphics[width=0.5\textwidth]{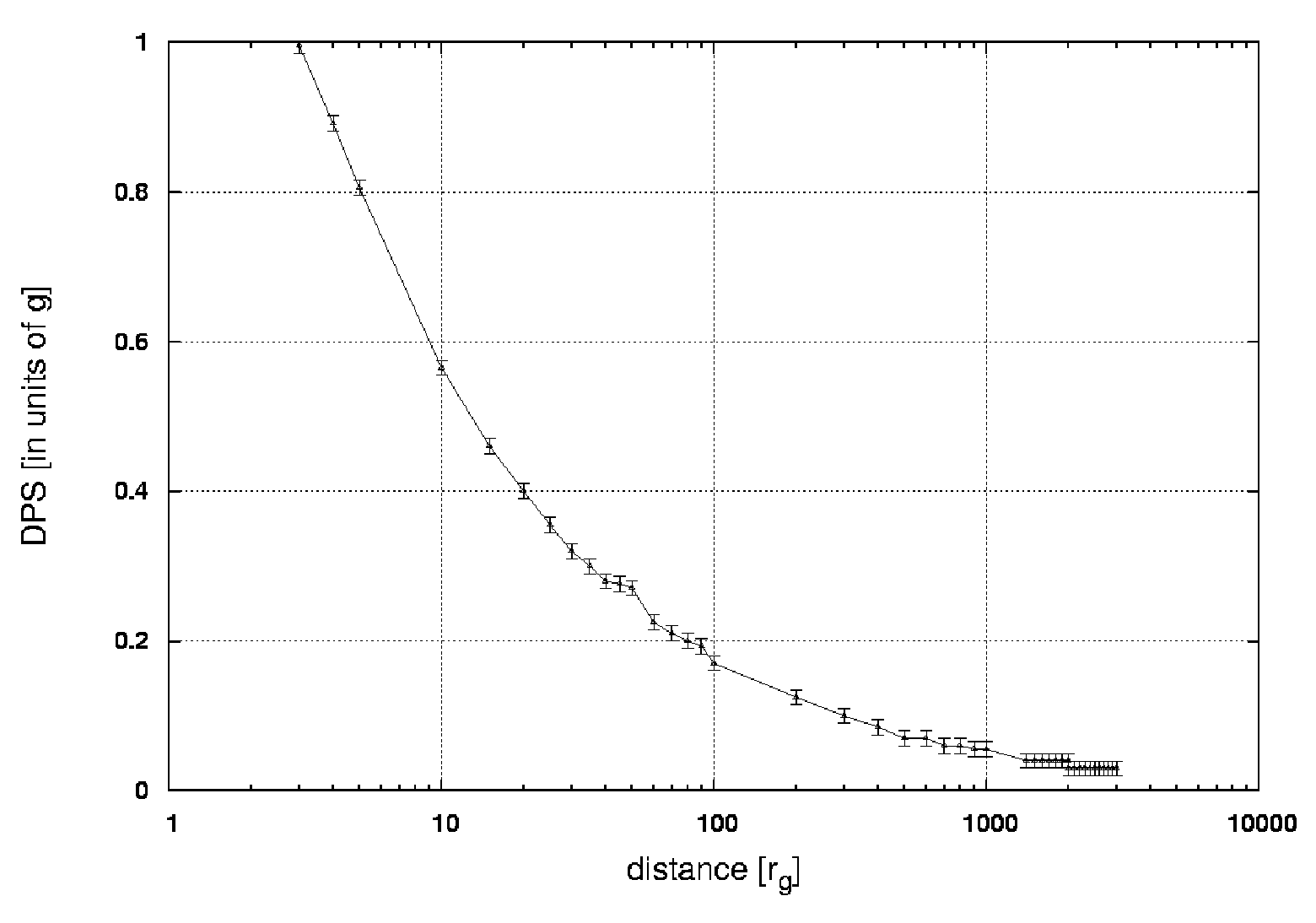}}
	\caption{Energetic distance of red and blue Doppler peak at $i=75^\circ$.
	This Doppler peak spacing (DPS) is measured in units of $g$. Far away from the black 
	hole both peaks approach significantly and the double--peaked line profile becomes 
	very narrow: At $\sim 3000 \ \mathrm{r_g}$ the peak difference in $g$ only amounts 
	to $\sim 0.03$. } \label{fig:DPS-i75}
\efigc
%%%%%%%%%%%%%%%%%%%%%%%%%%%%%%%%%%%%%%%%%%%%%%%%%%%%%%%%%%%%%%%%%%%%%%%%%%%%%%%%%%%%%%%%%%%%%%%%%%
%  
\section{Gravitationally redshifted optical emission lines} \label{sec:gropt}
\subsection{Observations of NLS--1 Mrk~110} \label{sec:mrk}
% cp. to Kollatschny 2003:
In a recent work by \cite{Kollatschny2003}, broad optical emission lines from 
H$\alpha$, H$\beta$, HeI$\lambda$5876 and HeII$\lambda$4686 were found to 
display significant and systematic redshifts. By reverberation mapping, the 
distances of the emitting regions from the central continuum source were
determined and a stratified ionization structure seen with HeII arising 
closest to the black hole at a distance of $\simeq 490 \ r_\mathrm{g}$. 
The observed shift of the HeII line was measured to be $\Delta z\simeq 0.002$, 
corresponding to $g\simeq 0.998$.
\subsection{Inner disk inclination of Mrk~110 from Kerr ray tracing} \label{sec:inc}
In this section, we follow the assumption that the observed optical lines of Mrk~110 
are subject of gravitational redshift and Doppler shifts and compare them with Kerr 
ray 
tracing techniques. It is aimed to determine parameters of the black hole--BLR 
system. The flat BLR model assuming line emitting rings as outlined in Sec. 
\ref{sec:param} is applied. Line redshifts are computed with the fully generalized 
GR Doppler factors $g$ that were presented by M04 (see Eq. 13 therein). We note 
that in order to capture the correct $g_\mathrm{core}$ value, a fine line binning 
is needed and hence a very high numerical spectral resolutions is used, 
$\Delta g = 0.00001$.

In contrast to our redshift analysis we now have to allow for arbitrary inclination 
angles for the BLR in Mrk~110. A parameter space with inclination between $1^{\circ}$ 
and $40^{\circ}$ and radii in the range 100 and 10000 $r_\mathrm{g}$ suffices to cover 
the observational data. The simulated core redshift values follow a power law, $z = p\,r^s$, 
for fixed inclinations and are plotted in Fig. \ref{fig:CpKollat1}. Observational data 
for Mrk~110 by K03 are overplotted as boxes. The horizontal error bars are due to the
uncertainty in time lag measurements and the vertical error bars involve the uncertainties
in the differential redshifts of the rms line centers. The labels of the axes are adapted 
to theoretical considerations and display redshift $z$ as a function of distance to the 
black hole in units of gravitational radii. To rescale the x--axis the best-fit value 
for the black hole mass\footnote{In general, it is recommended to treat $M$ as free 
parameter, too.} obtained by K03 is assumed, $M\simeq 1.4\times10^8\sol$. 
\bt
\caption{Power law fit parameters $p$ and $s$ for inclinations $1^{\circ}$ 
to $40^{\circ}$}
\label{tab:02}
\centering
\btu{ccccc}
\hline
\hline
  &  &  & & \\
$i\,[^{\circ}]$ & $p$ & error $\Delta p$ & $s$ & error $\Delta s$ \\
  &  &  & & \\
\hline
\hline
1 & 1.566 & $\pm$ 0.007 & -1.001 & $\pm$ 0.001 \\
10 & 1.428 & $\pm$ 0.008 & -0.999 & $\pm$ 0.001 \\
20 & 1.187 & $\pm$ 0.003 & -0.996 & $\pm$ 0.001 \\
30 & 0.914 & $\pm$ 0.013 & -1.007 & $\pm$ 0.003 \\
40 & 0.484 & $\pm$ 0.013 & -1.006 & $\pm$ 0.005 \\
\hline
\hline
\etu
\et
Tab. \ref{tab:02} shows the fitting results for power laws at each inclination 
angle. At any inclination the power law exhibits the same slope, 
$s_\mathrm{num}\approx -1.002\pm0.005$, which is the average of the five 
inclination angles assumed here. This is a direct consequence of the Schwarzschild 
factor, $z(r)\propto r^{-1}$ i.e.\ $s_\mathrm{theo}=-1$. Additionally, the power 
laws shift toward lower gravitational redshifts as the inclination angle increases. 
This is due to projection effects -- included in the \textit{projection parameter} $p$. 
The inclination dependence of the redshift can be approximated by $z\propto\cos(i)$
i.e.\ $p$ is related to the cosine of $i$. In other words, the higher the inclination, 
the more blueshift there will be (see Fig. \ref{fig:gcore-i1vs75}), and this Doppler
blueshift counteracts the redshift. Interestingly, this behaviour could be exploited 
to determine the inclination angle of the inner disk from observed gravitationally 
redshifted features. 

The best--fitting power law is shown as a thick solid line 
in Fig. \ref{fig:CpKollat1}.  
%displays the best fitting power law for the K03 data 
Here, the slope of $s=-1$ has been fixed and the projection parameter $p$ fitted to 
give $p_\mathrm{Mrk\,110}\simeq 0.886\pm 0.034$. Its relation to a specific inclination 
$i$ can be extracted from Fig. \ref{fig:CpKollat2} which shows the cosine behaviour of 
$p$ for the simulated sets $i\in\left\{1,10,20,30,40^{\circ}\right\}$. The cosine fit 
to ray tracing data yields $p(i)\simeq 4.63\,\cos(i)-3.13$ (error $\sim 10\%$; note that 
this is only valid for $i<40^{\circ}$) within the radial range between 100 and 10000 
$r_\mathrm{g}$. Taking the best fit value 0.886 one reads in Fig. \ref{fig:CpKollat2} 
at the cross that the inner disk of Mrk~110 is inclined to $i\simeq 30^{\circ}$. This 
result is consistent with that of K03 ($i\simeq 21\pm 5^{\circ}$).
%%%%%%%%%%%%%%%%%%%%%%%%%%%%%%%%%%%%%%%%%%%%%%%%%%%%%%%%%%%%%%%%%%%%%%%%%%%%%%%%%%%%%%%%%%%%%%%%%%
%
% Fig. 7: Kollatschny data in comparison with redshift factor from Kerr ray tracing simulations
%         for different inclination angles [in theoretical units]
\bfigc
	\rotatebox{0}{\includegraphics[width=0.5\textwidth]{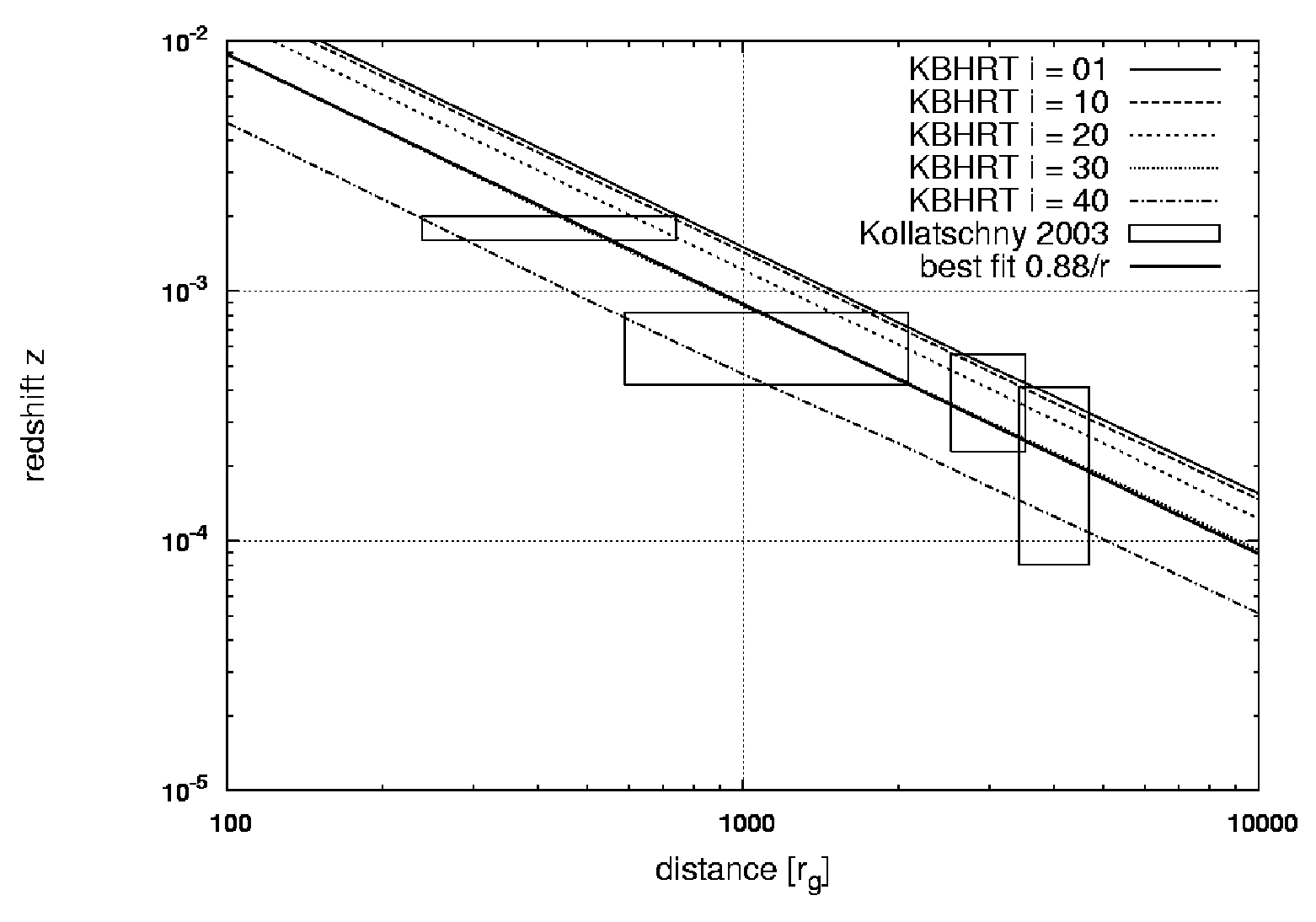}}
	\caption{Radial dependence of the line core redshift: data for the NLS--1 Mrk~110 
	taken from K03 (\textit{boxes}) are compared to the line core redshifts as computed from Kerr 
	ray tracing simulations with different inclination angles. Redshift as a function of distance 
	scales with a power law with identical slope $s=-1$ for all inclinations but with different 
	projection parameter $p$ that determines the vertical shift of the power law. Best 
	fit for Mrk~110 (\textit{solid thick}) yields $p\simeq 0.886$ that can be used to 
	determine the inclination of the inner disk.} \label{fig:CpKollat1}
\efigc
%%%%%%%%%%%%%%%%%%%%%%%%%%%%%%%%%%%%%%%%%%%%%%%%%%%%%%%%%%%%%%%%%%%%%%%%%%%%%%%%%%%%%%%%%%%%%%%%%%
%
%%%%%%%%%%%%%%%%%%%%%%%%%%%%%%%%%%%%%%%%%%%%%%%%%%%%%%%%%%%%%%%%%%%%%%%%%%%%%%%%%%%%%%%%%%%%%%%%%%
%
% Fig. 8: Variation of the projection parameter with inclination and determination of
%         inner disk inclination angle in Mrk 110 based on Kollatschny data
\bfigc
	\rotatebox{0}{\includegraphics[width=0.5\textwidth]{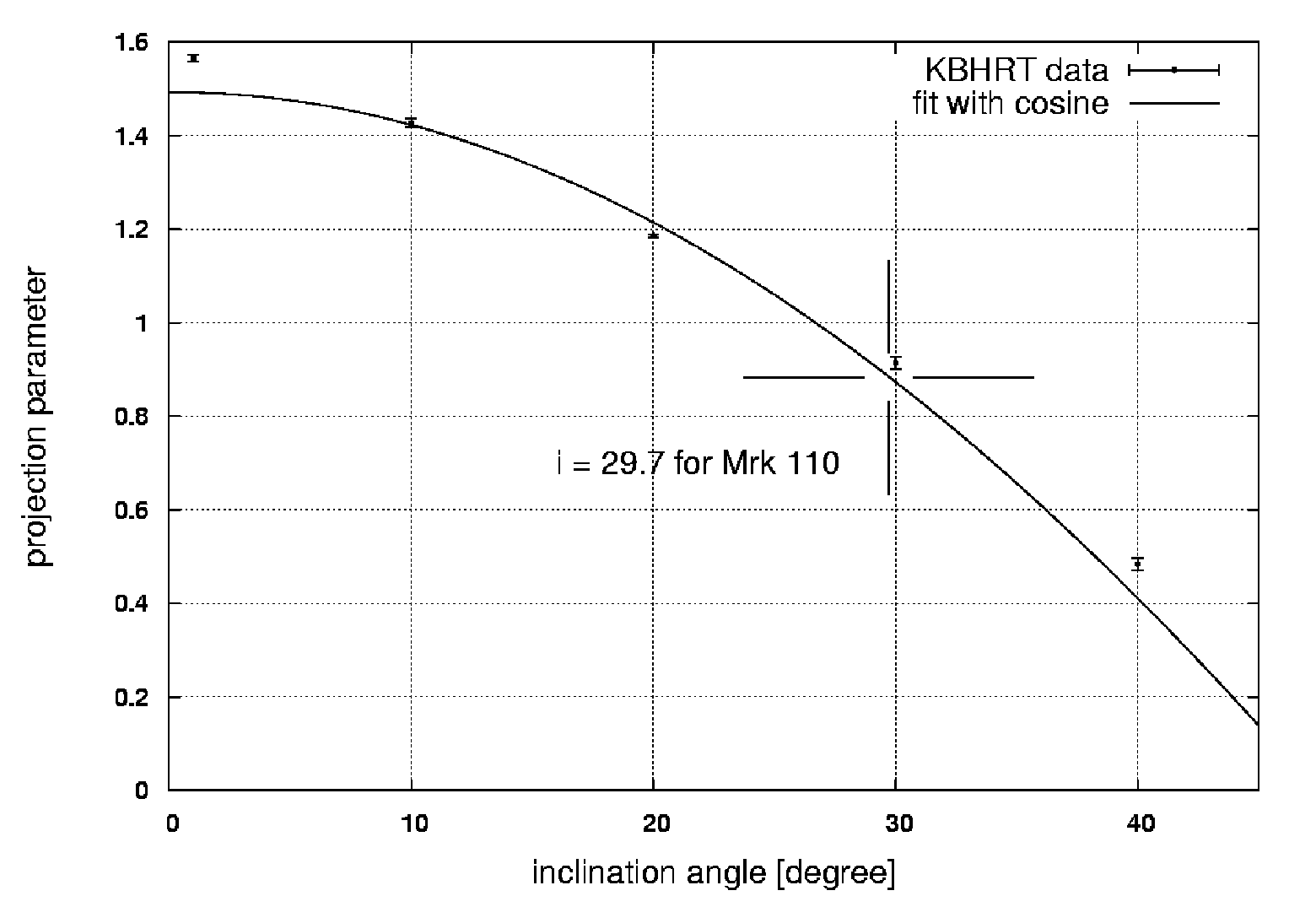}}
	\caption{Variation of the projection parameter $p$ with inclination $i$ as computed from Kerr 
	ray tracing. The numerical data points can be approximated by a cosine function (\textit{solid}), 
	$p(i)\sim 4.63\,\cos(i)-3.13$ within 10 \% precision. $p$ measurements from observations 
	give inner disk inclinations. The cross marks the result for Mrk~110 based on K03 data: 
	$i\sim (29.7\pm 1.2)^{\circ}$.} \label{fig:CpKollat2}
\efigc
%%%%%%%%%%%%%%%%%%%%%%%%%%%%%%%%%%%%%%%%%%%%%%%%%%%%%%%%%%%%%%%%%%%%%%%%%%%%%%%%%%%%%%%%%%%%%%%%%%
%
\subsection{Static vs. stationary emitter velocity field} \label{sec:vel}
% next: velocity field, static vs. stationary, lapse in Schw+Kerr
The redshifts calculated above were based on a simple, but \textit{dynamical} BLR model. 
In this section we clarify under which conditions the Schwarzschild factor can be used 
to estimate gravitational redshifts. Usually, a velocity field of the BLR emitters has 
to be assumed. The prominent Schwarzschild factor can only be applied to quantify the 
redshift in case of \textit{static} emitters (and static black holes). Of course, real 
emitters such as the BLR are dynamical e.g. in stationary motion. In this case, the ray 
tracing technique is a convenient method for redshift computations. This approach is also 
justified by the fact that it takes into account that real emitters are extended. The 
relativistic $g$--factor (consult M04 for details) for static emitters simplifies 
significantly because the Keplerian angular velocity satisfies $\Omega_\mathrm{K}=0$ for 
radii larger than the radius of marginal stability, $r>r_\mathrm{ms}$. The emitter velocity 
field in Bardeen Observer's frame is assumed to be purely rotational i.e.\ $v^{(i)}=0$ with 
$i=r,\,\theta$. We are thus left with a Lorentz factor $\gamma=\left(1-v_{(\Phi)}v^{(\Phi)}\right)^{-1/2}$. 
Considering $\Omega=\Omega_\mathrm{K}=0$ in $v^{(\Phi)}=\tilde\omega\,\left(\frac{\Omega-\omega}{\alpha}\right)$ 
the Lorentz factor simply goes to unity, $\gamma\rightarrow 1$. It should be kept in mind 
that this regime is \textit{not} applicable at $r<r_\mathrm{ms}$ when frame--dragging $\omega$ 
becomes important and $\Omega=0$ is replaced by $\Omega\rightarrow\omega$. Altogether, the 
generally complicated $g$--factor is identical to the lapse function $\alpha$ for static 
emitters in the $r>r_\mathrm{ms}$ regime. The quantity $\alpha$ is evaluated here for static 
black holes (as already shown in K03) and for rotating black holes. Restriction to the 
equatorial plane, $\theta=\pi/2$, yields a simple expression for the lapse function (see 
e.g. M04, Eq. 3):
\beq
\alpha=\sqrt{\frac{r(r^2-2Mr+a^2)}{r^3+a^2(r+2M)}}.
\eeq
The lapse function reduces to the well--known Schwarzschild factor for $a=0$
\beq
\alpha\mid_{a=0}\,=\sqrt{1-\frac{2M}{r}}.
\eeq
\subsection{Black hole rotation} \label{sec:rot}
A \textit{rotating} SMBH can not be excluded a priori in the case of Mrk~110. On the 
contrary, theories of black hole growth, see e.g.\ \cite{Shapiro2005}, strongly suggest 
fast spinning SMBHs in the local universe with either $a/M\simeq 0.95$ (MHD disk) or 
even $a/M\simeq 1$ (standard thin gas disk). However, it is also known that the rotation 
of space--time decays very rapidly in the outer regions of the black hole gravitational 
potential. The frame--dragging frequency decreases according to $\omega\propto r^{-3}$. 
This is documented in the final Fig. \ref{fig:CpKollat3} which summarizes the results 
from Fig. \ref{fig:CpKollat1} and \ref{fig:CpKollat2} and also extends beyond the region 
explored by K03. The redshifts corresponding to $g$--factors of $g\longmapsto\alpha$ are 
plotted for a non--rotating and a fast rotating black hole with $a=0$ (\textit{solid curve}) 
and $a/M=0.998$ (\textit{dotted curve}), respectively. The curves are compared to the 
observations of Mrk~110 (\textit{boxes}) and the best-fitting $i=30^{\circ}$ Kerr ray 
tracing simulation (\textit{filled circles}). 
%
%%%%%%%%%%%%%%%%%%%%%%%%%%%%%%%%%%%%%%%%%%%%%%%%%%%%%%%%%%%%%%%%%%%%%%%%%%%%%%%%%%%%%%%%%%%%%%%%%%
%
% Fig. 9: Kollatschny data in comparison with best fitting Kerr ray tracing simulation with i = 30
%         and in comparison with lapse functions in Schwarzschild/Kerr [in theoretical units]
\bfigc
	\rotatebox{0}{\includegraphics[width=0.5\textwidth]{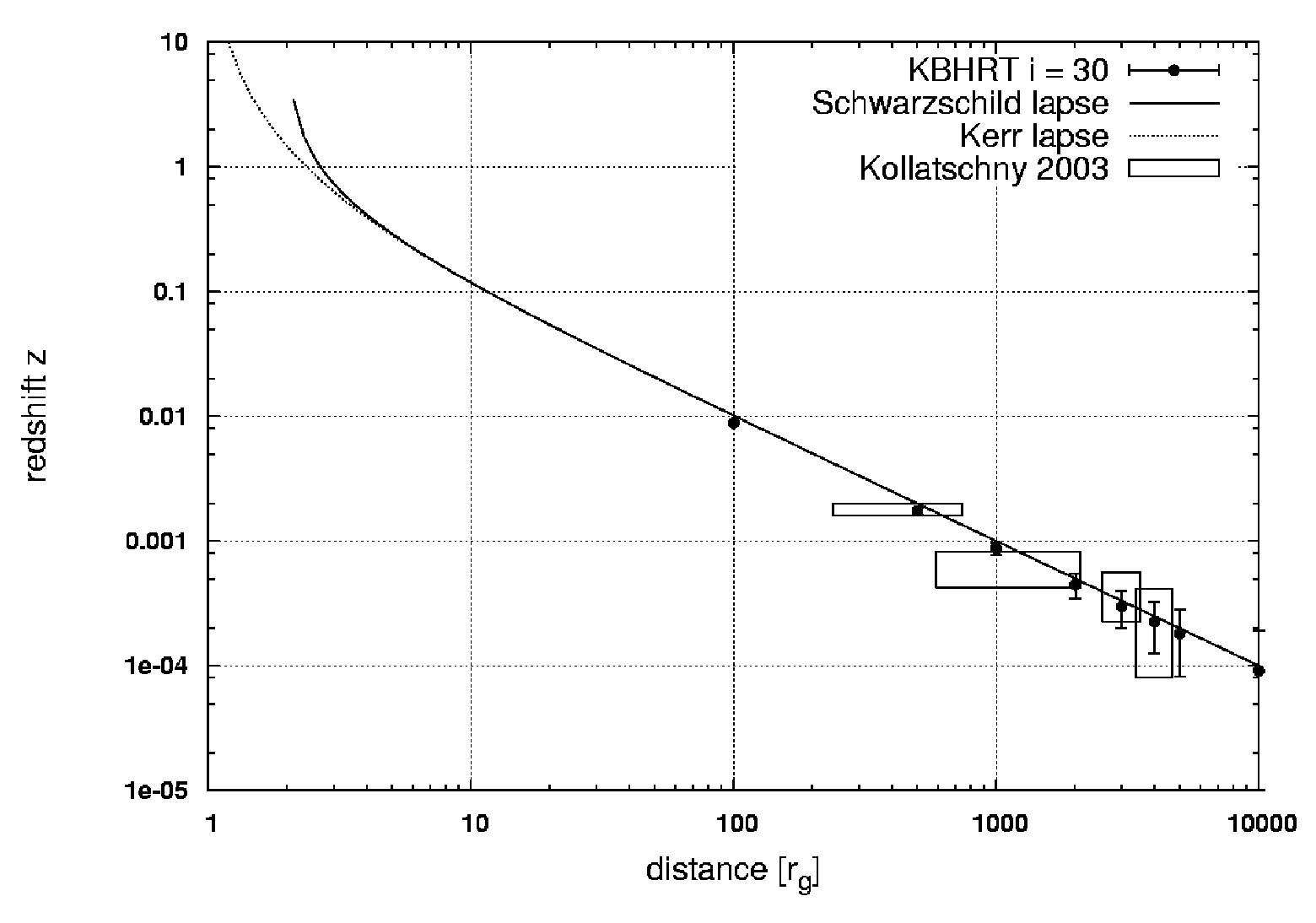}}
	\caption{Synoptical plot with optical K03 data for Mrk~110 (\textit{boxes}) and best fitting
	Kerr ray tracing simulation with $i\sim 30^{\circ}$ (\textit{filled circles}) as well as lapse 
	functions for a static (Schwarzschild, \textit{solid}) and rotating (Kerr, \textit{dotted}) 
	black hole. An essential statement illustrated here is that optical BLR lines can \textbf{not} 
	probe black hole rotation due to their huge distance. Generally, multi--wavelength observations
	are recommended to fill the gap at smaller radii.} \label{fig:CpKollat3}
\efigc
%%%%%%%%%%%%%%%%%%%%%%%%%%%%%%%%%%%%%%%%%%%%%%%%%%%%%%%%%%%%%%%%%%%%%%%%%%%%%%%%%%%%%%%%%%%%%%%%%%
%
It is evident from the figure that the rotation of space--time is important only at small radii, 
$r\lesssim 4 \ \mathrm{r_g}$. Therefore, black hole rotation can only be probed with spectral features 
originating in regions very close to the black hole, like the X--ray fluorescence lines of iron. 
Optical emission lines are not suited for probing black hole rotation in that manner -- at least not
for AGN. The reason for the small offset between the ray tracing results shown by the filled circles 
in Fig. \ref{fig:CpKollat3} and the Schwarzschild/Kerr lapse functions plotted as solid and dotted 
lines is the different velocity field of the emitters and the transverse Doppler effect. The 
ray traced emitters are assumed to rotate stationarily, whereas the lapse functions only coincide with 
ray traced $g$--values when static emitters are assumed. Also the ray traced redshift values are based 
on $g_\mathrm{core}$ computed by averaging over the whole line (Eq.~\ref{eq:gcore}), and therefore are 
less sharp than those given by the lapse function.

Multi--wavelength observations such as the ongoing COSMOS project\footnote{\texttt{http://www.astro.caltech.edu/$\sim$cosmos/}} 
may help filling the gap between 2 and 200 $\mathrm{r_g}$ in Fig. \ref{fig:CpKollat3}. If several 
gravitationally redshifted spectral lines can be identified in a source, the central mass can be 
determined with high accuracy. 
%Considering one source many spectral features
%in different spectral ranges are subject of systematic gravitational redshifts from the central mass.
%If a huge amount of high--quality multi--wavelength data is available the central mass can be pinpointed 
%with high--accuracy. 
%
\subsection{Assumptions} \label{sec:crit}
% new paragraph 
BLR emission lines are influenced by both gravitational redshift and Doppler
effects. Therefore, the fitting procedure involves two parameters, black 
hole mass $M$ and inclination angle of the emitter $i$. The inclination 
can be determined if the black hole mass is known from other methods. In 
case of Mrk~110, the mass of the SMBH was inferred from reverberation 
mapping as well as from gravitational redshift in the literature. 
% It is easier to determine 
% masses based on reverberation mapping. However, in this case there remains 
% an uncertainty due to the unknown orientation of the emitter. If the 
% signal--to--noise and the resolution are good enough, gravitational redshift 
% helps to get an alternative estimate of the black hole mass. 
Other techniques are measurements involving stellar and gas dynamics, the 
$M$--$\sigma$ relation and Masers. 
To be able to determine the inner inclination, we used the best fit 
black hole mass computed from gravitational redshift found in K03.
% main finding 
The error boxes of K03 can be fitted quite well with the ray tracing 
results (3.9~\% error). Hence, a model with stationary rotating rings 
distributed in a radial range between 240 and 4700 $\mathrm{r_g}$ and 
inclined to $i\sim 30^{\circ}$ can explain the observed gravitational 
redshift of optical lines in Mrk~110.
% H$\alpha$, H$\beta$, HeI$\lambda$5876 and HeII$\lambda$4686 in Mrk 110. 
%An AGN type--1 in Mrk 110 gets additional support from these studies. 
%type-1s are defined according to the width of their broad emission lines,
%not according to the inclination of an assumed BLR.
It is surprising that a simple Keplerian rotating model can describe the 
BLR structure so well. Similar results have been obtained e.g.\ for NGC 
5548 \citep{Peterson1999} and other AGN \citep{Peterson2000, Onken2002, Chen1989}.  
%The elliptical galaxy Arp 102B is another example that showed evidence for a 
%relativistic Keplerian disk \citep{Chen1989}.
In this work we have made the assumptions of a flat and Keplerian rotating
BLR structure. However, nature is supposed to be more complicated allowing 
particularly for BLR wind components and opacity effects 
\citep{Popovic1995, Chiang1996}. This implies a modification of the solver
including radial and poloidal motion as well as photon propagation through 
a dense volume.
Furthermore, disk warping has not been considered and this may significantly 
change the results. For warped disks the analysis has to be extended by using 
suitable generalized ray tracing codes e.g. like the one by \cite{Cadez2003}.
\section{Conclusions} \label{sec:conc}
Line cores at distances from 2 to 100000 $r_\mathrm{g}$ from a rotating black hole 
have been analysed using relativistic ray tracing simulations in the Kerr geometry. 
The line cores are gravitationally redshifted by 
$z_\mathrm{core}\simeq \ 10^{-5},\,10^{-4},\,10^{-3},\,10^{-2},\,10^{-1}\,10^{0}$ at 
distances of $100000,\,10000,\,1000,\,100,\,10,\,2 \ \mathrm{r_g}$ from the black hole, 
respectively. This $z\approx\frac{1}{\mathrm{distance[r_g]}}$ behaviour at large radii 
is a straightforward consequence of the Schwarzschild factor. Lines characterised by 
a core energy $g_\mathrm{core}$ confirm this scaling behaviour. 
%The gravitational redshift effect has correspondingly $0.002,\,0.02,\,0.2,\,1.5,\,15,\,60$\% significance. 

Gravitational redshift occurs in two modes. One regime starts at larger distances from 
the black hole and shifts only the line as a total feature while conserving its intrinsic 
shape; the amount of redshift can be looked up in Tab. \ref{tab:01}. The other regime, 
which is the strong gravity regime, dominates 
%Another regime termed the strong gravity dominates in the 
%innermost black hole surroundings 
at $r\lesssim 10 \ \mathrm{r_g}$ or at $r \lesssim 10^{13}$ cm for a ten million solar 
mass black hole. Relativistic emission lines originating in this region are strongly 
deformed and suppressed and differ substantially from the corresponding line profiles 
in the emitter's rest frame. 

The present work has demonstrated that the onset of GR becomes important at 
distances smaller than $75000 \ \mathrm{r_g}$ if assuming an optical spectral resolution 
of 0.1 \AA. It is stressed that this critical radius depends on the astronomical resolving 
power and lies farther out if the resolution is higher. However, even with high--resolution 
spectroscopy it remains a challenge to probe gravitationally redshifted spectral lines due 
to the fact that competing effects and more complex physics are likely to be involved: a 
flat Keplerian BLR model may be too simple to model the complex BLR velocity field and both 
spherical BLR structure and a wind component may play a crucial role. 
In addition, narrow-line components and e.g.\ contamination of the H$\beta$ line by FeII 
may also complicate both the analysis and the interpretation. Nevertheless, the theory of 
General Relativity predicts that the gravitational redshift is present, and a valuable 
ansatz for probing it is to search for systematic shifts with varying distance, as done by 
K03. It is suggested in this work to supplement this by multi--wavelength observations that 
should all point towards the same central mass and inner inclination. 

We confirm the analysis by K03 of the NLS--1 galaxy Mrk~110 here, using a more general 
treatment with stationary $g$--factors that are in concordance with observational data.
Ray tracing simulations in the Kerr geometry support an inclination of $i=30^{\circ}$ for 
the inner disk of Mrk~110. Reversely, if the inclination is known, this can be exploited to 
determine the black hole mass from the fitting procedure outlined here. Whether fitting $i$ 
or $M$ (or both) -- such techniques may help explore AGN unification schemes: multi--wavelength 
studies allow for studying the inclination deep into the AGN and to probe orientation and 
luminosity--dependence of AGN types. Furthermore, we show that broad optical lines can not 
serve as a probe of black hole spin because frame--dragging effects only occur very close to 
the black hole. In this region, only hot emission lines (such as the Fe K$\alpha$ line in 
X--rays) or other relativistic spectral features indicate black hole spin. The analysis 
presented here is not only valid for supermassive black holes but also for stellar--mass 
black holes in X--ray binaries or intermediate--mass black holes that may be found in 
ultra--luminous X--ray sources or globular clusters.
%
% BIB
%
%
\begin{acknowledgements}
AM wishes to thank the organizers and participants of the Japanese--German meeting in Wildbad 
Kreuth, Germany, especially Wolfram Kollatschny (University of G\"ottingen) and Lutz Wisotzki 
(AIP). We wish to thank John Silverman (MPE) for inspiring discussions.
\end{acknowledgements}
\bibliographystyle{aa}
\bibliography{GRlinesbib}
\end{document}